\documentclass[conference]{IEEEtran}
\ifCLASSINFOpdf
  \usepackage[pdftex]{graphicx}
\else
\fi

\PassOptionsToPackage{hyphens}{url}
\usepackage[hidelinks]{hyperref}

\usepackage{booktabs}
\usepackage{listings}
\usepackage{multirow}
\usepackage{amssymb}
\usepackage{pifont}
\usepackage[table]{xcolor}
\usepackage[]{algorithm2e}
\let\labelindent\relax
\usepackage{enumitem}
\usepackage[keeplastbox]{flushend}

\usepackage{tikz}

\begin{document}

\title{Intrinsic Rowhammer PUFs:\\ Leveraging the Rowhammer Effect \\for Improved Security\vspace{-17.5pt}}


%

\author{\IEEEauthorblockN{Andr\'{e} Schaller\IEEEauthorrefmark{1},
Wenjie Xiong\IEEEauthorrefmark{2},
Nikolaos Athanasios Anagnostopoulos\IEEEauthorrefmark{1},
Muhammad Umair Saleem\IEEEauthorrefmark{1},\\
Sebastian Gabmeyer\IEEEauthorrefmark{1},
Stefan Katzenbeisser\IEEEauthorrefmark{1} and
Jakub Szefer\IEEEauthorrefmark{2}}
\IEEEauthorblockA{\IEEEauthorrefmark{1}Technische Universit{\"a}t Darmstadt and CYSEC, Darmstadt, Germany} \IEEEauthorblockA{\IEEEauthorrefmark{2}Yale University, New Haven, CT, USA}
\vspace{1.5pt}
}

\newcommand\copyrighttext{%
  \footnotesize This document is based on the accepted version of the work of the same title that has been published in the \href{https://ieeexplore.ieee.org/xpl/mostRecentIssue.jsp?punumber=7940186}{Proceedings of the 2017 IEEE International Symposium on Hardware Oriented Security and Trust (HOST)} by IEEE, which has the following Digital Object Identifier (DOI): \href{https://doi.org/10.1109/HST.2017.7951729}{10.1109/HST.2017.7951729}. Minor formatting changes have been applied.\\
  \\
\textcopyright 2017 IEEE. Personal use of this material is permitted. Permission from IEEE must be obtained for all other uses, in any current or future media, including reprinting/republishing this material for advertising or promotional purposes, creating new collective works, for resale or redistribution to servers or lists, or reuse of any copyrighted component of this work in other works.}

\newcommand\copyrightnotice{%
\begin{tikzpicture}[remember picture,overlay]
\node[anchor=south,yshift=10pt, xshift=5pt] at (current page.south) {\fbox{\parbox{\dimexpr\textwidth-2\fboxsep-2\fboxrule\relax}{\copyrighttext}}};
\end{tikzpicture}%
}

\maketitle

\begin{abstract}

Physically Unclonable Functions (PUFs) have become an important and promising hardware primitive for device fingerprinting, device identification, or key storage. Intrinsic PUFs leverage components already found in existing devices, unlike extrinsic silicon PUFs, which are based on customized circuits that involve modification of hardware. In this work, we present a new type of a memory-based intrinsic PUF, which leverages the Rowhammer effect in DRAM modules -- the {\em Rowhammer PUF}. Our PUF makes use of bit flips, which occur in DRAM cells due to rapid and repeated access of DRAM rows. Prior research has mainly focused on Rowhammer attacks, where the Rowhammer effect is used to illegitimately alter data stored in memory, e.g., to change page table entries or enable privilege escalation attacks. Meanwhile, this is the first work to use the Rowhammer effect in a positive context -- to design a novel PUF. We extensively evaluate the Rowhammer PUF using commercial, off-the-shelf devices, not relying on custom hardware or an FPGA-based setup. The evaluation shows that the Rowhammer PUF holds required properties needed for the envisioned security applications, and could be deployed today.

\end{abstract}




%

\copyrightnotice

\enlargethispage*{-20pt}
\vspace{-15pt}
\section{Introduction}

In recent years, attacks that exploit the Rowhammer effect have gained a lot of attention, as they can enable a plethora of security-related risks due to the wide-spread vulnerability imposed by the Rowhammer effect in today's DRAM modules. The phenomenon was first described by Kim et al.~\cite{kim2014flipping}, who were able to induce so-called disturbance errors in high-density, commodity DRAM modules by repeatedly accessing uncached memory rows. Disturbance errors occur due to the charge coupling between DRAM cells, which accelerates charge leakage in adjacent rows, and eventually results in bits being flipped in so-called victim rows in DRAM, even though said victim rows were not explicitly accessed. The Rowhammer effect allows for breaking many software-based security mechanisms, as well as memory and process isolation, because it allows flipping memory bits, which would otherwise be protected by software-based access control mechanisms. Numerous papers have been published that use the Rowhammer effect in order to improve the identification of vulnerable DRAM cells or to implement various Rowhammer attacks~\cite{seaborn2015exploiting, veen:2016,xiao2016one,razavi2016flip}.

In contrast to the existing work on the Rowhammer effect, we present a novel approach that uses DRAM disturbance errors, in order to strengthen the security of DRAM-equipped devices, instead of attacking such platforms. We propose to use bit flips, induced by the Rowhammer effect, as basis for a Physically Unclonable Function (PUF) that allows for robust identification of DRAM-equipped devices. We further present a software-only solution that works on commodity hardware and which enables runtime queries to the Rowhammer PUF, not requiring custom hardware or an FPGA setup. Prior work on DRAM PUFs has considered using the decay characteristics of DRAM cells when refresh is disabled, e.g.~\cite{Xiong:2016}, but the Rowhammer effect as a source of a PUF has not been explored so far. Compared to existing DRAM decay-based PUFs, the Rowhammer PUF takes advantage of disturbance errors to increase the entropy of the PUF response. With our new approach, we enable DRAM-equipped low-cost platforms to use hardware-based fingerprinting, identification, or key storage mechanisms without adding extra logic, e.g., as opposed to extrinsic arbiter PUFs that require new circuits to be added to the computing platform. Since many, if not most, DRAM-equipped platforms are affected by the Rowhammer effect~\cite{kim2014flipping}, application of Rowhammer PUF goes well beyond just the platforms tested in this work. Additionally, unlike most known intrinsic PUFs, particularly SRAM-based PUFs, which can only be accessed at SRAM boot-up time, the Rowhammer PUF can be queried both at boot-up time and at runtime of a system.

\subsection*{Contributions}

This paper extends the field of Physically Unclonable Functions (PUFs) with the following contributions:
\renewcommand\labelitemi{\tiny$\bullet$}
\begin{itemize}[leftmargin=*]
\setlength{\itemsep}{0pt}
\item We introduce the Rowhammer PUF, which leverages disturbance errors among DRAM rows that manifest themselves as bit flips, which are used as basis for the new type of Physically Unclonable Function.
\item We implement the Rowhammer PUF on commodity, off-the-shelf devices, in a way which is accessible during runtime and which requires no custom hardware or an FPGA setup.
\item We provide an extensive evaluation, showing very good metrics for uniqueness, robustness and entropy. We further show the PUF's ability to operate at different ambient temperatures in a stable manner.
\end{itemize}
\enlargethispage*{-20pt}
\section{Background}
\label{sec:prelim}

\subsection{DRAM Data Storage and Access}
\label{sec:prelim:dram}

\begin{figure}[t]
\centering
\includegraphics[width=3.5in]{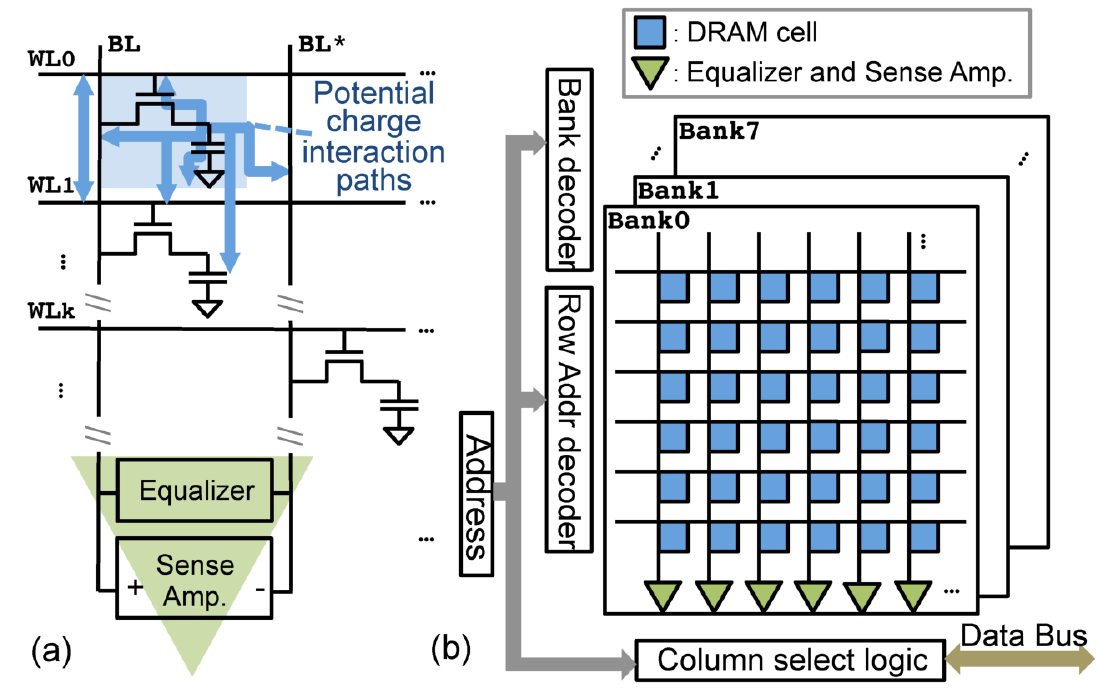}
\caption{(a) Schematic of DRAM cells: the blue arrows show potential charge interaction paths. (b) DRAM organization.}
\label{fig:DRAM}
\end{figure}

DRAM stores a ``bit'' as charge on a capacitor. A single DRAM cell consists of a capacitor for storage and a transistor for access, as shown in Figure~\ref{fig:DRAM}(a). The gate of the transistor is connected to a {\em wordline} (WL). Each wordline controls access to the whole row. The capacitor is connected to a {\em bitline} (BL) through the transistor. Each bitline is also connected to an {\em equalizer} and a {\em sense-amplifier} to convert charge on the capacitor to a digital signal. DRAM cells are organized in arrays, as in Figure~\ref{fig:DRAM}(b). Each array is also called a bank. Usually, a DRAM chip consists of 8 banks.

The charge on the capacitor will leak over time, and data in the cell will be lost. Figure~\ref{fig:DRAM}(a) shows several charge interaction paths where the charge may leak. The time until a cell loses its data is called the {\em data retention time}. To keep data for longer time, wordlines of each row must be accessed periodically, so that the sense-amplifiers recharge the capacitors of that row through the bitlines. This process is called ``DRAM refresh''. To ensure data integrity, every DRAM row needs to be refreshed with a certain frequency, which is usually $32$ to $64$ms in current DRAMs.

\subsection{The Rowhammer Effect in DRAM}
\label{sec:prelim:rowhammer}

The Rowhammer effect, which is based on disturbance errors in DRAM cells, has been discovered in recent years~\cite{kim2014flipping}. It is an unintended side effect in DRAM that occurs when one memory row (the {\em hammer row}) is rapidly and repeatedly accessed. This causes cells in nearby rows to leak charge more quickly and thereby introduces changes (i.e., ``bit flips'') to the contents of the affected memory cells. This is due to the interaction between adjacent wordlines as well as between DRAM cells and their neighbouring capacitors and wires, as depicted in Figure~\ref{fig:DRAM}(a). It has been shown that hammering a row will most likely affect its two adjacent rows. Consequently, {\em double-sided} Rowhammer, where both adjacent rows of a victim row are hammered, has been proposed to increase the chance of bit flips~\cite{seaborn2015exploiting}.

Usually, to allow for a sufficiently high DRAM access rate, and thus to trigger disturbance errors, non-cached memory accesses are needed, e.g., by leveraging the $\mathtt{CLFLUSH}$ instruction. Lately, several works have demonstrated the feasibility to exploit the Rowhammer effect on platforms that do not provide such cache line flush instructions. In order to circumvent CPU caching mechanisms and ensure direct access to DRAM, Gruss et al.~\cite{gruss2015rowhammer} and Aweke et al.~\cite{aweke2016anvil} enforce cache eviction through elaborate memory access patterns. Qiao and Seaborn~\cite{Qiao2016} make use of x86 non-temporal store instructions, which do not use the CPU cache and van der Veen et al.~\cite{veen:2016} utilize non-cacheable DMA queries to exploit the Rowhammer effect. Other papers have presented techniques to gain understanding of the locations of flipping bits. Razavi et al.~\cite{razavi2016flip} presented a technique that allows for targeted bit flips at arbitrary physical memory locations by combining the Rowhammer effect with memory duplication. In order to conduct predictable Rowhammer attacks, van der Veen et al.~\cite{veen:2016} use a brute-force approach to hammer all DRAM rows and collect information about expectable bit flip locations.

Since its discovery, the Rowhammer effect has been used mainly as means of attacking a computer system. In particular, changing the contents of memory cells can result in modification of important data. Seaborn and Dullien~\cite{seaborn2015exploiting} as well as van der Veen et al.~\cite{veen:2016} rely on the Rowhammer effect in order to gain root privileges, by flipping bits in page table entries. Xiao et al.~\cite{xiao2016one} attack Xen's paravirtualized memory isolation by employing the Rowhammer effect from within a malicious virtual machine. Razavi et al.~\cite{razavi2016flip} as well as Bhattacharya and Mukhopadhyay~\cite{bhattacharya2016curious} successfully attack RSA by creating bit flips in keys stored in DRAM.

Meanwhile, to the best of our knowledge, this is the first work that leverages the Rowhammer effect in a positive way.

\subsection{Memory-Based Intrinsic PUFs}
\label{sec:prelim:pufs}

In this work we focus on the class of intrinsic PUFs, which do not require the addition of extra circuits to a device in order to use them. Unlike extrinsic PUFs that necessitate addition of circuitry, e.g., arbiter circuits, intrinsic PUFs only use standard hardware components already present in commodity computer systems, such as SRAM or DRAM memory arrays. SRAM PUFs have been studied extensively and are based on the startup values that SRAM cells take on after powering on a device~\cite{guajardo2007fpga,schrijen2012comparative}. DRAM-based PUFs have so far only leveraged DRAM cells' start-up values~\cite{tehranipoor2015dram,7579621}, or DRAM cell-decay effects~\cite{Xiong:2016}. In contrast, this work presents the Rowhammer PUF, which is a new type of an intrinsic, DRAM-based PUF that uses the Rowhammer effect.

\section{Rowhammer PUF in Commodity DRAM}
\label{sec:impl}

Previous work has shown that the locations of disturbance errors in DRAM cells are stable~\cite{kim2014flipping,veen:2016}. This makes the Rowhammer effect a promising candidate for a PUF. However, the number of bit flips introduced by the Rowhammer effect can be relatively small, and thus may only provide a limited amount of entropy. We thus introduce three techniques to help increase entropy, without changing the physical DRAM properties. First, we disable DRAM refresh for those memory locations where the PUF is located. This prevents the PUF cells from being recharged, as would happen if normal refresh was on, and increases the number of bit flips. Second, multiple DRAM rows are used together to create an instance of the Rowhammer PUF that encompasses larger amount of cells. Third, hammering time and initial values of the DRAM cells are controlled to induce a maximum number of bit flips.

\subsection{Rowhammer PUF Parameters}
\label{sec:impl:parameters}

\begin{figure}[t]
\centering
\includegraphics[width=3.2in]{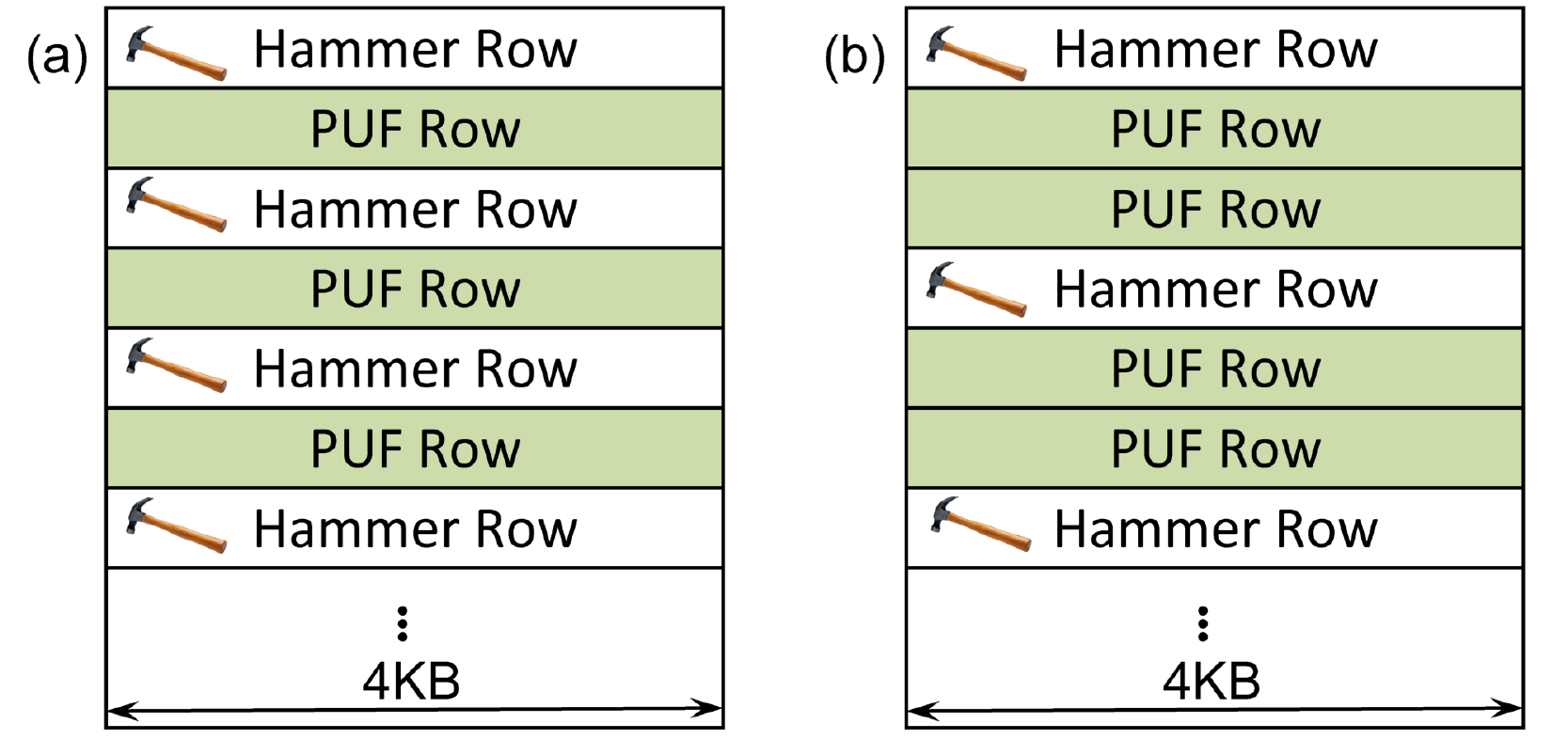}
\caption{Rowhammer types: (a) Double-sided Rowhammer (\texttt{DSRH}) with \texttt{PUF size}~$=12$KB; (b) Single-sided Rowhammer (\texttt{SSRH}) with \texttt{PUF size}~$=16$KB. We assume DRAM with $4$KB row size.}
\label{fig:hammer_pattern}
\end{figure}

There are many parameters that can influence the Rowhammer PUF. In Section~\ref{sec:eval}, we present an evaluation upon which the most suitable values among these parameter are selected.

\noindent\textbf{Rowhammer type:} As presented in the literature~\cite{seaborn2015exploiting,aweke2016anvil}, there are two approaches, or Rowhammer types (\texttt{RH types}), in order to induce the Rowhammer effect. If for one victim row there is only one adjacent hammer row, used to induce bit flips, we call it single-sided Rowhammer (\texttt{SSRH}). In contrast, double-sided Rowhammer (\texttt{DSRH}) involves exposing both neighbors of a particular victim row as hammer rows. The patterns of hammer and victim rows (also called PUF rows), used to conduct \texttt{SSRH} and \texttt{DSRH} are shown in Figure~\ref{fig:hammer_pattern}.

\noindent\textbf{PUF address and size:} The \texttt{PUF address} defines the starting address of a PUF in the DRAM. The \texttt{PUF size} depends on the the number of rows that are examined after the Rowhammer process finishes (PUF rows in Figure~\ref{fig:hammer_pattern}). Due to the existence of hammer rows, PUF rows are not consecutive. \texttt{PUF size} and \texttt{RH type} influence the actual hammering frequency, as smaller PUFs will allow each hammer row to be accessed more frequently. Likewise, \texttt{SSRH} has fewer hammer rows, so each can be accessed more often within the same time~period.

\noindent\textbf{Hammer~row~IV (initial value):}
For the memory range that corresponds to \texttt{PUF address} and \texttt{PUF size}, corresponding hammer rows will be pre-initialized by writing \texttt{Hammer row IV} to it, before conducting the Rowhammer process.

\noindent\textbf{PUF row IV (initial value):} Similarly, all PUF rows that are included in this memory range are initialized with \texttt{PUF row IV} before the Rowhammer process is started. Both, \texttt{Hammer row IV} and \texttt{PUF row IV} are important parameters because disturbance errors are caused by interaction of DRAM cell charges. Moreover, DRAM cells represent a particular logic value using different charge states, resulting in so-called true-cells and anti-cells~\cite{liu2013experimental}. True-cells represent a logic `1' as charge on the capacitor and `0' as no charge, while anti-cells do the opposite. Consequently, initializing a true-cell with `0' would not lead to a bit flip in that cell. Thus, it is important to evaluate the effect of different values of \texttt{PUF row IV} and \texttt{Hammer row IV}. As the layout of true- and anti-cells is identical for DRAM modules of the same type, once optimal settings for both initial values have been found, they can be re-used for other instances of the same device type.

\noindent\textbf{Rowhammer time:} The Rowhammer time (\texttt{RH time}) defines the total duration of the PUF measurement, including disabling the refresh rate and conducting the hammering process. \texttt{RH time}, just as \texttt{PUF size} and \texttt{RH type}, affects how many times each hammer row will be accessed in total.

\subsection{Rowhammer PUF Access}
\label{sec:impl:access}

Given the above parameters, the process of accessing a Rowhammer PUF is depicted in Algorithm~\ref{alg:rowhammerpuf}. Based on \texttt{PUF address}, \texttt{PUF size}, and \texttt{RH type}, the DRAM region for the Rowhammer PUF is defined. First, this DRAM region is reserved, such that no other program accesses the same region. Next, the PUF rows and hammer rows are initialized with \texttt{Hammer row IV} and \texttt{PUF row IV}, respectively. The PUF query is started by disabling the DRAM auto-refresh in the next step. This is done using the same technique as employed in~\cite{Xiong:2016}. Subsequently, the process of hammering respective rows is started. For this purpose, the hammer rows need to be accessed repeatedly for a certain time. This is achieved by a read operation to the first word of each hammer row, which in turn causes the whole DRAM row to be refreshed. Hence, bits in the PUF rows will start to leak charge and will eventually flip. After \texttt{RH time}, the process ends and the DRAM auto-refresh is enabled again. Finally, the PUF measurement can be read from the PUF rows.

\RestyleAlgo{boxruled}
\begin{algorithm}[t]
	\small
	\KwData{\texttt{RH\_type}, \texttt{PUF\_address}, \texttt{PUF\_size}, \texttt{Hammer\_row\_IV}, \texttt{PUF\_row\_IV}, \texttt{RH\_time}}
	\KwResult{PUF measurement $m$}
	$\cdot$ reserve memory defined by \texttt{PUF\_address} and \texttt{PUF\_size}\;
	$\cdot$ initialize \textit{PUF\_rows} with \texttt{PUF\_row\_IV} and \textit{hammer\_rows} with \texttt{Hammer\_row\_IV}\;
	$\cdot$ disable auto-refresh of PUF rows\;
	\While{$t<\mathtt{RH\_time}$}{
		\For{$r_i$ $\in$ hammer\_rows}{
			$\cdot$ read access to row $r_i$\;
		}
	}
	$\cdot$ enable auto-refresh\;
	$\cdot$ read \textit{PUF\_rows} as PUF measurement $m$\;

	\caption{\footnotesize Process of the Rowhammer PUF query.}
	\label{alg:rowhammerpuf}
\end{algorithm}

\subsection{Factors Affecting the Rowhammer PUF}
\label{sec:impl:factors}

Because the Rowhammer PUF is inherently tied to the underlying physical properties of the DRAM modules, there are three factors that can influence the operation of the PUF.

\noindent\textbf{Temperature:} Prior work has shown that Rowhammer victim cells are not strongly affected by temperature~\cite{kim2014flipping}. However, the Rowhammer PUF is based on the interaction of the Rowhammer effect and DRAM decay, which was shown to be temperature-sensitive. Thus, we evaluate the temperature effect in Section~\ref{sec:eval}, which confirms that the Rowhammer PUF exhibits increased bit flips but stable noise values at higher operating~temperatures.

\noindent\textbf{Voltage:} Prior work has also shown that voltage affects the leakage in DRAM cells \cite{hamamoto1998retention}. In commodity, off-the-shelf devices there is currently no interface to control the voltage of DRAM cells. We assume that for the Rowhammer PUF, the DRAM operates at the factory specified voltage parameters. Voltage factors will be investigated in future work.

\noindent\textbf{Error Correcting Codes (ECC):} ECC can be used in DRAM to protect from bit flips. Many computing platforms, such as the PandaBoard used in this work, do not have ECC implemented. Even if ECC is present, the authors of~\cite{aichinger2015ddr} showed that ECC is not enough to mitigate the Rowhammer effect. In order to use the Rowhammer PUF when ECC is used, the PUF size would have to be increased. Further, ECC registers that indicate rows, which observed bit flips, could potentially be exploited for PUF measurements. We will explore this in future work. In this paper we assume that no ECC is used.

\subsection{Software Implementations in Commodity Devices}
\label{sec:impl:implementation}

In this paper, the Rowhammer PUF is implemented and tested on the PandaBoard~\cite{PandaBoard}. The PandaBoard ES Revision B3 used in our experiments houses a TI OMAP 4460 System-on-Chip (SoC) module and 1GB DDR2 memory from ELPIDA in a Package-on-Package (PoP) configuration, which operates at 1.2V. Our PUF implementation is purely in software, leaving hardware configuration unchanged.

The Rowhammer PUF is implemented in the U-Boot boot loader. Since the DRAM is idle during U-Boot runtime, queries to the Rowhammer PUF can be conducted without affecting other functions of the platform. In U-Boot, one can control the DRAM refresh cycle. Further, one can access physical DRAM addresses without caching\footnote{PandaBoard implements an ARM processor that does not provide the \texttt{CLFLUSH} instruction. Thus, we avoid caching by querying the Rowhammer PUF during an early stage during DRAM initialization, before caching is enabled by the boot loader.}.

The reference manuals provide the physical address mapping of the DRAM. We allocate hammer rows and PUF rows in \texttt{Bank0} and make them adjacent, as shown in Figure~\ref{fig:hammer_pattern}(a--b).

It is further possible to access the Rowhammer PUF from within a kernel module to achieve runtime access. Similar to U-Boot, DRAM refresh can be disabled from kernel space. Moreover, in contrast to U-Boot, where caching can be avoided by accessing the Rowhammer PUF during an early point of DRAM initialization, the kernel module allows for disabling caching by setting respective register values can be disabled if the platform does not support the \texttt{CLFLUSH} instruction.


\section{Evaluation}
\label{sec:eval}

In this section, we will provide details on the various characteristics of the proposed Rowhammer PUF. We will first discuss how different values of the parameters presented in Section~\ref{sec:impl:parameters} affect the number of observed bit flips. We then evaluate the Rowhammer PUF on the basis of a fixed parameter configuration with regards to uniqueness, robustness and entropy. Finally, we discuss varying ambient temperature conditions that could influence the Rowhammer PUF.

We will follow an explorative approach, which involves assessment of a subset of all potential parameter values. Due to the lack of information about the distribution of true- and anti-cells\footnote{Usually this is the case when dealing with commercial, off-the-shelf devices as most vendors treat such implementation details regarding their hardware components as intellectual property and thus will not disclose them.}, it is necessary to explore the correlation between parameter values and PUF behavior experimentally, by testing various parameter settings.

In our evaluation, three different memory regions, each located on one individual PandaBoard, are measured\footnote{In the following we denote such a memory region as a \textit{PUF instance}.}. For all of the measurements, the PUF address was fixed. For each parameter combination, 20 measurements were taken.

\subsection{Effects of Rowhammer Parameters}
\label{sec:eval:parameters}

Given the high dimensional parameter space (see Section~\ref{sec:impl:parameters}), we focussed on evaluating such configuration settings that are expected to yield a good PUF. An overview of all evaluated parameter settings is given in Table~\ref{tab:parameterspace}. For the sake of brevity, we only present most important results of the extensive evaluation data we obtained.

\begin{table}[!t]
	\caption{Parameters used for evaluation of the Rowhammer PUF characteristics, and their corresponding set of values.}
	\label{tab:parameterspace}
	\centering
	\begin{tabular}{lc}
		\toprule
		Parameter	&	Evaluated Values	\\
		\midrule
		\texttt{RH type}	&	single-sided (\texttt{SSRH}), double-sided (\texttt{DSRH})	\\
		\texttt{PUF size}		&	$4$KB, $32$KB, $128$KB	\\
		\texttt{Hammer row IV}	&	`\texttt{0x00}', `\texttt{0x55}', `\texttt{0xAA}', `\texttt{0xFF}'	\\
		\texttt{PUF row IV}	& `\texttt{0x00}', `\texttt{0x55}', `\texttt{0xAA}', `\texttt{0xFF}'	\\
		\texttt{RH time}	&	$60$s, $120$s	\\
		\bottomrule
	\end{tabular}%
\end{table}

In order to extract the maximum possible entropy from PUF measurements, we primarily strive to maximize the number of bit flips. For this purpose, we first identify those parameters that have the largest effects on the amount of bit flips. In the following, we will discuss the parameters listed in Table~\ref{tab:parameterspace}, in the context of their impact on observable bit flips.

\begin{table*}[!t]
	\caption{Overview of the average number of observable bit flips, depending on combinations of \texttt{Hammer row IV} and \texttt{PUF row IV}. Configuration used: \texttt{PUF size}$=128$KB and \texttt{RH time}$=120$s (\texttt{SSRH}/\texttt{DSRH}).}
	\label{tab:rowIVsettings}
	\centering
	\begin{tabular}{cc|c|c|c}
		\toprule
			 & \multicolumn{4}{c}{Hammer row IV}	\\
			 \cline{2-5}
			 & & & & \\
			PUF row IV & `\texttt{0x00}' & `\texttt{0x55}' & `\texttt{0xAA}' & `\texttt{0xFF}'	\\
			\midrule
			'\texttt{0x00}' & $7405$ / $8032$ & $17558$ / $20358$ & $7391$ / $7200$ & $17288$ / $20152$	\\
			'\texttt{0x55}' & $0$ / $0$ & $0$ / $0$ & $0$ / $0$ & $0$ / $0$	\\
			'\texttt{0xAA}' & $22547$ / $24480$ & $\mathbf{32904}$ / $\mathbf{37548}$ & $14218$ / $14243$ & $24479$ / $28268$ 	\\
			'\texttt{0xFF}' & $15633$ / $17798$ & $15402$ / $17579$ & $6132$ / $6479$ & $6095$ / $6416$	\\
		\bottomrule
	\end{tabular}%
\end{table*}

\begin{figure*}[t]
\vspace{0.5cm}
\begin{minipage}[t]{.5\textwidth}
  \centering
(a)~\includegraphics[height=2.25in]{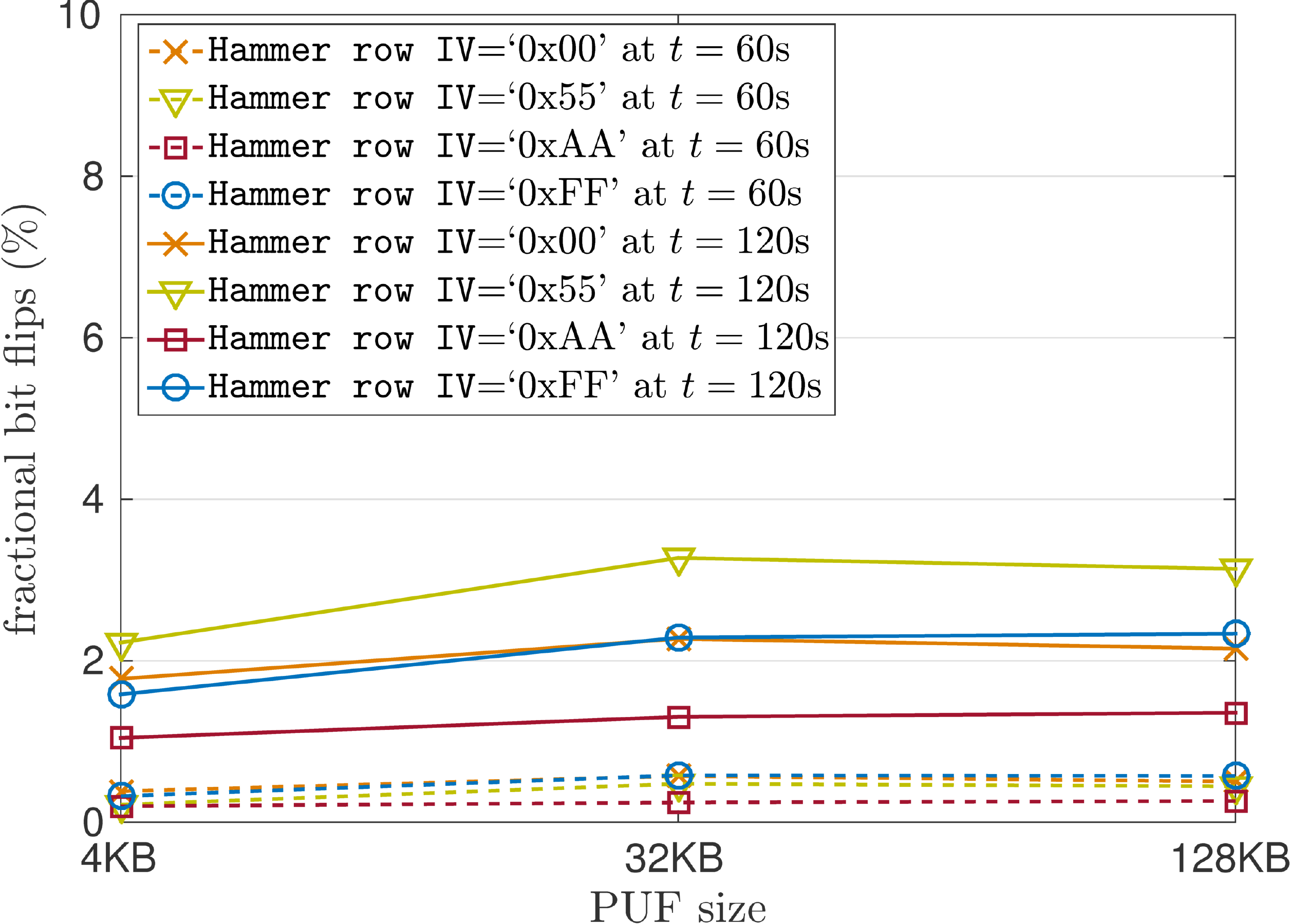}
\end{minipage}%
\begin{minipage}[t]{.5\textwidth}
  \centering
(b)~\includegraphics[height=2.25in]{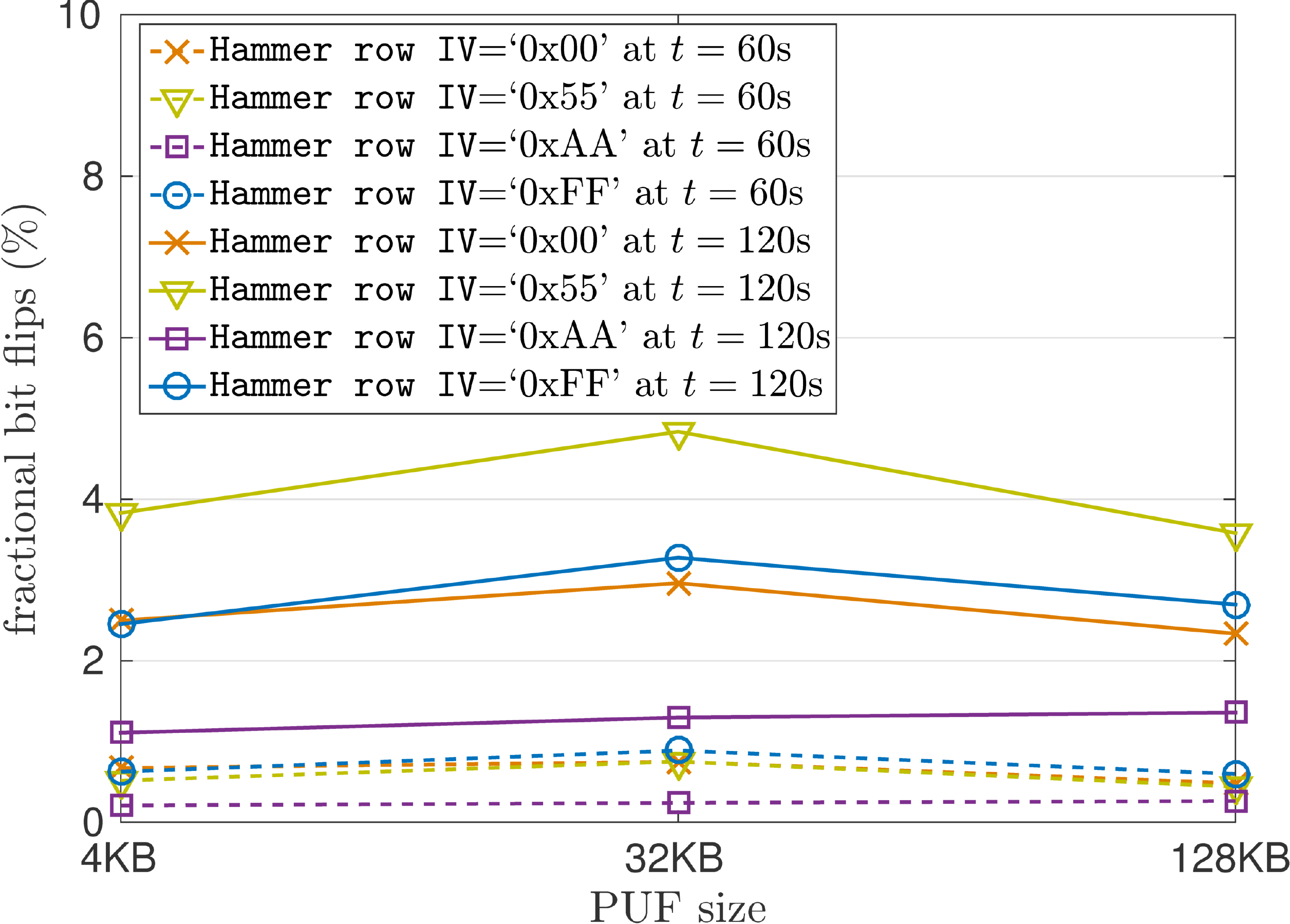}
\end{minipage}
\caption{Fractional number of bit flips, given in percent relative to \texttt{PUF size}, using \texttt{PUF
row IV}$=$`\texttt{0xAA}'. Left: number of bit flips using \texttt{SSRH}. Right: number of bit flips using \texttt{DSRH}.}
\label{fig:rhtypes}
\end{figure*}

The bit flips we observe in the Rowhammer PUF measurements are due to the hammering process and the DRAM cell decay that emerges after DRAM refresh is disabled. In order to confirm that the Rowhammer effect adds a significant number of extra flips, we measured the number of bit flips, which are solely caused by the decay process, and compared it to the total number of bit flips we observed in the Rowhammer PUF measurements. Compared to the bit flips caused by DRAM decay, the Rowhammer PUF introduces $2.4$ times bit flips in $60$ seconds and about twice the number of bit flips in $120$ seconds. Further, the set of bits that flip (i.e., their locations in the measurements) obtained from the Rowhammer PUF only partially overlaps with the set of bit flips induced by the DRAM decay process, even for longer decay times (i.e., without the influence of the Rowhammer effect). Thus, the Rowhammer PUF induces new bit flips, which are at different locations compared to the DRAM decay process.

\noindent\textbf{Rowhammer type:}
Initially, we expected the \texttt{RH type} parameter to have a strong influence on the number of flipped bits. In Figures~\ref{fig:rhtypes}(a--b), we present the \textit{fractional} number of bit flips as a percentage of the absolute number of bits available (\texttt{PUF size}). Contrary to our expectations, applying \texttt{DSRH} (right) instead of \texttt{SSRH} (left), does not lead to a highly increased number of flips, despite hammering both rows adjacent to respective PUF rows. Instead, compared to \texttt{SSRH}, using \texttt{DSHR} only leads to $\approx9\%$ more bit flips in $60$ seconds and to $\approx15\%$ in $120$ seconds on average.

\noindent\textbf{PUF size:}
The \texttt{PUF size} influences the total time required to execute a single iteration of hammering the DRAM. In our implementation, each hammer row is accessed roughly every $6\mu s$ when hammering 2 rows ($4$KB PUF) and every $8\mu s$ when hammering 17 rows ($128$KB PUF) in the \texttt{SSRH} setting. Figures~\ref{fig:rhtypes}(a--b) show the number of bit flips relative to \texttt{PUF size}. The number of bit flips does not change significantly for different values of \texttt{PUF size}, i.e., the fraction of bit flips for different memory ranges stays stable.

\noindent\textbf{Hammer row and PUF row IV:}
Given that DRAM arrays consist of true-cells and anti-cells, the initial value (IV) of the hammer rows as well as the PUF rows is expected to play an important role regarding the number of observable bit flips. Depending on the type of a cell, a bit flip in a PUF row can be observed only if the cell is initialized with the logic value that corresponds to its charged state. Similarly, due to physical interaction of charged analog elements in the hammer and PUF rows (i.e., wires and capacitors) and the resulting charge interaction paths, different IVs of the hammer rows influence the number of bit flips as well. Thus, the values of both parameters must be chosen carefully, in order to maximize bit flips. As can be seen from Table~\ref{tab:rowIVsettings}, different configurations of \texttt{Hammer row IV} and \texttt{PUF row IV} lead to measurements that exhibit different bit flips. In general, it can be inferred from the experiments, that the number of bit flips on the PandaBoard can be maximized, if PUF rows are pre-initialized in such a way that keeps true-cells and anti-cells in the charged state, whereas cells of the adjacent hammer rows are kept in an uncharged state. In particular, the measurements show that most bit flips can be observed, if PUF rows are initialized with `\texttt{0xAA}', which depicts a bit-wise checkerboard pattern, with a leading `1'. Adjacent hammer rows are set up using the complementary pattern, starting with a `0' bit (`\texttt{0x55}'). In contrast, no bit flips can be observed when initializing PUF rows with '\texttt{0x55}', as in this case, cells of the PUF rows were initialized corresponding to their uncharged states.\\


\noindent\textbf{Summary:}  The parameters \texttt{Hammer row IV} and \texttt{PUF row IV} have a predominant influence on the number of bit flips, which we strive to maximize. We therefore first fix their values as follows: \texttt{Hammer row IV}$=$`\texttt{0x55}' and \texttt{PUF row IV}$=$`\texttt{0xAA}'. We further set \texttt{RH type} to \texttt{SSRH}, as the number of introduced bit flips is in the same order of magnitude as for \texttt{DSRH}. Additionally, \texttt{SSRH} requires $\approx55\%$ less memory and involves less memory accesses compared to \texttt{DSRH}. Although setting \texttt{RH time}$=120$s leads to $\approx400\%$ more bit flips as for $60$s, we will provide evaluation results for both parameter values. The resulting PUF readout time is similar to existing runtime accessible decay-based DRAM PUFs.

\subsection{PUF Characteristics}
\label{sec:eval:PUFmetrics}

In order to assess the applicability of the set of flipped bits as a PUF, we validated uniqueness, robustness and entropy of the Rowhammer PUF measurements, using the parameter configuration identified above. Instead of using metrics that are based on the Hamming distance (i.e., inter- and intra-Hamming distance), we follow the approach of~\cite{Xiong:2016} and utilize the \textit{Jaccard index}~\cite{jaccard1901etude}. This is motivated by the fact that DRAM-based PUFs show different characteristics compared to classic PUFs (such as SRAM-based PUFs). In particular, measurements from DRAM modules draw their PUF characteristics from the location of the flipped bits. This fact (uniqueness of indices) is not properly reflected in Hamming distance-based measures. Let $s_x$ denote the set of indices of flipped bits in the corresponding PUF measurement $m_x$. The Jaccard index between two measurements is calculated as $J(s_1,s_2)=\frac{\left|s_1\cap s_2\right|}{\left|s_1\cup s_2\right|}$, depicting their similarity.

\noindent\textbf{Uniqueness and Robustness:}
Uniqueness is measured by means of the $J_{inter}$ metric. This metric compares the indices of bit flips between pairs of measurements obtained from different PUF instances. Ideally, for maximal PUF uniqueness, the two sets should have no common elements (i.e., no overlaps of bit flip locations), resulting in $J_{inter}$ values close to `0'. In contrast, the $J_{intra}$ metric measures the PUF's robustness, in particular the influence of noise, present in subsequent PUF measurements obtained from a fixed PUF instance. Here, the elements of two sets should be identical in the best case, resulting in $J_{intra}$ close to `1'.

\begin{figure}[!t]
\centering
\includegraphics[width=3.05in]{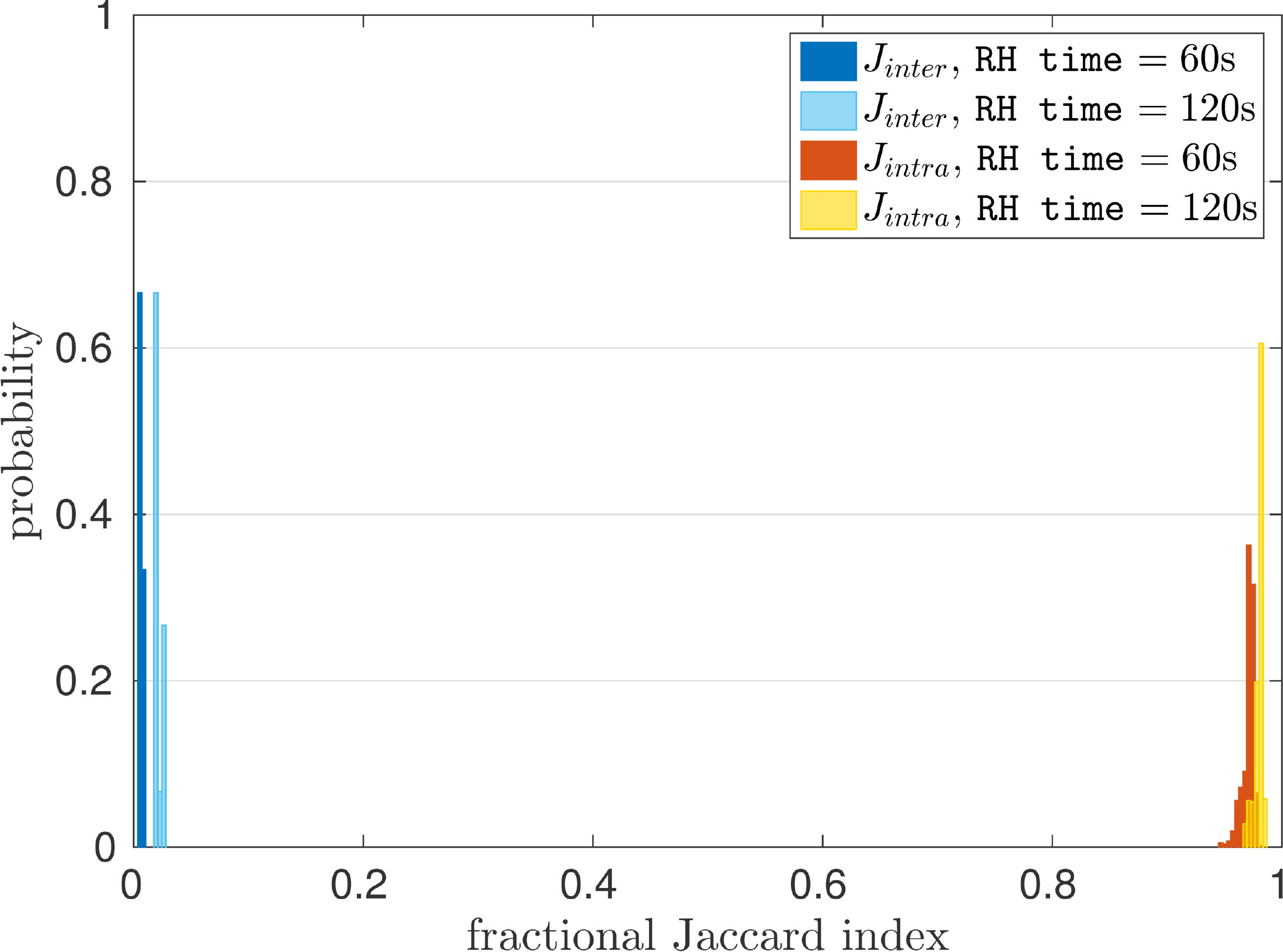}
\caption{Histogram of $J_{inter}$ and $J_{intra}$ values for three PUF instances using 20 measurements with \texttt{PUF row IV}$=$`\texttt{0xAA}', \texttt{Hammer row IV}$=$`\texttt{0x55}', \texttt{PUF size}~$=128$KB and \texttt{RH type} set to \texttt{SSRH}.}
\label{fig:jaccard}
\end{figure}

Figure~\ref{fig:jaccard} shows histograms of values obtained for both metrics. Clearly, both histograms separate, indicating that Rowhammer PUF instances can be robustly and uniquely identified. With a minimum $J_{intra}$ value of $0.9454$, Rowhammer PUF measurements exhibit a maximum noise of $\approx5\%$, which can be easily corrected by standard Fuzzy Extractor constructions~\cite{dodis2004fuzzy}.

\noindent\textbf{Entropy:} PUF measurements should exhibit sufficient entropy in order to derive a cryptographic key. We estimated the entropy of the PUF measurements, as proposed in~\cite{Xiong:2016}. Denoting the total number of bits contained in a PUF measurement (i.e., \texttt{PUF size}) as $N$, $k$ as the cardinality of $s_x$ and assuming that the locations of flipped bits are distributed uniformly, the entropy can be calculated as follows: $H=log_2{{N}\choose{k}}$.

Using the minimum number of bit flips ($k=30994$) observed in the measurements, based on the optimized parameter setting identified above ($N=1048576~\hat{=}~128$KB), the lower bound for the fractional entropy (i.e., the entropy per cell), is $0.192$. Given the vast amount of available cells, the PUF measurements show sufficient entropy to derive cryptographic keys. For example, the derivation of a $128$bit key, given entropy of $0.192$ requires approximately $85$Bytes (excluding entropy required to compensate leakage due to helper data), while the PUF is already $128$KB.

\noindent\textbf{Temperature Dependency:}
The behavior of DRAM bit flips can be influenced by the operating temperature. In order to validate usage of the Rowhammer PUF in several operating temperatures, we evaluated the PUF at temperatures of $40^\circ$C (working temperature of DRAM on Pandaboard), $50^\circ$C and $60^\circ$C. We computed the number of bit flips and $J_{intra}$ values for measurements taken at these respective temperatures. Table~\ref{tab:temperature} shows the average number of bit flips as well as the minimum $J_{intra}$ (i.e., maximum noise). The temperature evaluation shows that, while bit flips increase at higher temperatures, the noise level stays constant for each temperature. Thus, the Rowhammer PUF exhibits sufficient stability to be used at higher temperatures.

\begin{table}[!t]
	\caption{Comparison of the average number of bit flips and minimum $J_{intra}$ values obtained at operating temperatures of $40^\circ$C, $50^\circ$C and $60^\circ$C using the optimal parameter configuration for \texttt{PUF size} $=128$KB.}
	\label{tab:temperature}
	\centering
	\begin{tabular}{lccc}
		\toprule
		\multirow{2}{*}{Metric} & \multicolumn{3}{c}{Operational Temperature} \\
		\cline{2-4}
			& & & \\
			&$40^\circ$C & $50^\circ$C & $60^\circ$C \\
		\midrule
		avg. bit flips & 32904 & 65431 & 132450 \\
		min. $J_{intra}$ & 0.9662 & 0.9810 & 0.9847 \\
		\bottomrule
	\end{tabular}%
\end{table}
\section{Conclusion}
\label{sec:conclusion}

This paper presented the {\em Rowhammer PUF}, a new type of an intrinsic, memory-based PUF. Unlike the majority of work that has used the Rowhammer effect to trigger a security exploit, this is the first paper that uses the Rowhammer effect in a positive context. We extensively evaluated the Rowhammer PUF using commercial, off-the-shelf devices, not requiring custom hardware or an FPGA-based setup. The evaluation showed that the Rowhammer PUF holds required properties needed for device authentication or cryptographic key storage.

As with any new hardware security primitive, further work is required to expand the understanding of the Rowhammer PUF. Especially, we expect to perform studies investigating how voltage affects the Rowhammer PUF. Moreover, aging experiments with use of a thermal chamber need to be conducted to understand the long-term stability of this new PUF under various conditions. Finally, we will investigate how to improve the PUF readout time. The code of our design will be made available as Open Source at \url{http://caslab.csl.yale.edu/code/rowhammerpuf/}.

\section*{Acknowledgement}

This work has been partly funded by the German Research Foundation (Deutsche ForschungsGemeinschaft -- DFG), as part of project P3 within the CRC 1119 CROSSING, and, also, partly by the German Academic Exchange Service (Deutscher Akademischer AustauschDienst -- DAAD). This work has also been partly supported and funded by the US National Science Foundation (NSF) under NSF Grant no. 1651945.



\bibliographystyle{./IEEEtran}
\bibliography{./IEEEabrv,./IEEEexample}
%

\end{document}